\documentclass[preprint,sort&compress]{elsarticle}
\usepackage{natbib}
\usepackage{pifont}
\usepackage{geometry}
\usepackage{fleqn}
\usepackage{graphicx}
\usepackage{txfonts}
\usepackage{amssymb}
\usepackage{subfig}
\usepackage{color}
\usepackage{ulem}
\usepackage{float}
\usepackage{hyperref}
\usepackage{miller}

\begin{document}

\begin{frontmatter}
\title{Molecular dynamics simulations of the defect evolution in tungsten 
on successive collision cascades}

\author[barc]{Utkarsh Bhardwaj}
\ead{haptork@gmail.com}

\author[barc,Hbni]{M Warrier}
\ead{manoj.warrier@gmail.com}
        
\address[barc]{Computational Analysis Division, BARC, Visakhapatnam, 
Andhra Pradesh, India -- 530012}
        
\address[Hbni]{Homi Bhabha National Institute, Anushaktinagar, Mumbai,
Maharashtra, India -- 400094}

\begin{abstract}
Molecular dynamics (MD) simulations of successive collision cascades
within the same simulation domain were performed using two different 
inter-atomic potentials (IAP) in tungsten, one EAM based and the other a 
`quantum accurate' machine learning potential, SNAP. The micro-structural
changes are analyzed as a function of displacements per atom (dpa) for primary
knock-on atom (PKA) energies of 20 keV and 50 keV, reaching up-to irradiation
dose of 0.1 and 0.2 dpa, respectively. Five sample simulations are carried out
for each case for observing stochastic differences in the evolution of damage. A detailed
defect analysis is carried out to observe changes in different parameters such
as the number of surface defects, defect density, defect 
morphology and size distribution etc., as a function of dpa. We explore the
properties that are sensitive to the IAP used and those that are sensitive to
the PKA energy and note their similarities with experimental results at various
dpa values. The SNAP potential shows better agreement with the experiments for
swelling and number of surface defects. However, it also predicts presence of
high number of small sessile defects which may have definite affect on the
processes of microstructural evolution and material
properties.
\end{abstract}


\end{frontmatter}

\section{Introduction} \label{Introduction}

Irradiation induced changes in materials have been studied at multiple scales 
for several decades (\cite{PhysRev.120.1229, Zinkle201265, KaiHistRev2019, 
MFRT-KMC-2008} and references therein). The main focus of these studies have 
been on understanding the change in micro-structure of the materials due to 
irradiation. Primary damage at the atomistic scales due to collision cascades 
has been extensively studied (\cite{Stoller2012293, BACON20001, 
MALERBA200628}). The number of defects, their in-cascade clustering, defect 
morphologies, cascade properties, etc., are being studied from MD collision 
cascade databases \cite{CascadesDB, DEBACKER2021152887, BhardwajClassify, 
BhardwajSavi, BhardwajJOSS, BhardwajSubcascades}. More recently the effect of 
IAP used on the primary damage is being explored \cite{EffectZBLEDisp, 
PhysRevB.66.134104, MALERBA200628, TERENTYEV200665, MS, Bhardwaj3IAPCmp}. The 
evolution of the micro-structure is being studied using kinetic Monte Carlo 
(KMC) methods \cite{DomainOKMC2004, BECQUART201039} and mean field rate 
theoretical (MFRT) methods \cite{MFRT-KMC-2008}. The KMC methods 
use approximations for the migration energies of diffusion and morphological 
transitions, and interactions of defects with different morphologies and sizes.
KMC methods are also limited in their scale since only small regions of the 
order of microns can be simulated up-to small doses (fraction of a dpa) 
\cite{DomainOKMC2004}. The MFRT methods average over the spatial and temporal 
variations of the defect distribution and is less exact than the KMC methods, 
but can handle large regions for high doses \cite{MFRT-KMC-2008}. It is 
important to study the stability of the defect clusters 
\cite{Bhardwaj100Stability, LiuC15} their transport, their interactions with 
each other in order to understand the evolution of the micro-structure due to
irradiation. \\

Another approach used to quantify the irradiation exposure of materials is to 
define a unit that is physically related to the damage mechanism called 
"displacements per atom" (dpa) \cite{standard1994e693}. dpa contains 
information about both the fluence and the energy of the irradiating particle 
and is therefore a much better measure than the fluence (fig.1 of 
\cite{WasRadDamIonBeams}). A simple formula, called the NRT-DPA, based on the 
available energy per primary knock-on atom (PKA) and accounting for the 
displacement efficiency, electronic stopping and the displacement energy 
required to create a Frenkel Pair, was developed as a standard for evaluating 
dpa \cite{NORGETT197550}. The NRT-DPA, which is the current standard for 
quantifying energetic particle damage to materials for the past four decades, 
has several shortcomings like (i) over-estimating the number of Frenkel pairs 
produced and (ii) under-estimating the number of atoms replaced in the lattice 
\cite{KaiNatureComm2018}. Therefore newer, more accurate models, have been 
proposed \cite{INDC-NDS-0624, INDC-NDS-0648, KaiNatureComm2018}. \\

The studies at multiple scales mentioned in the first paragraph have yielded 
insights into irradiation induced damage. However, a 
lot more work on the stability, transport and interactions of defects have to 
be carried out using MD to be able to model micro-structure evolution due to 
irradiation accurately with higher scale models. A recent approach to straddle
over this shortcoming is based on the fact that the material properties show a
good correlation with dpa. Several researchers have created dpa at the atomic
level by a combination of random Monte Carlo moves of atoms followed by a
conjugate gradient relaxation of the atomic positions called the creation
relaxation algorithm (CRA) \cite{CRAZrDudarev,CRAHinWMason}. Mason et al. have
got better results by introducing a molecular dynamics collision cascade
cascade in-between several Monte Carlo steps \cite{CRAHinWMason}. Byggm\"astar
et al., have studied the effect of IAP on the evolution of defects at high
doses by carrying out successive MD simulations of collision cascades in W at 10 keV
\cite{EffectZBLEDisp}. They have validated their results by using
machine-learning (ML) interatomic potential for relaxation of simulation cells
that are obtained using non-ML interatomic potentials. The ML interatomic
potential is considered quantum accurate but relatively slow to carry out
high-dose irradiation simulations.\\

The experiments using techniques such as TEM \cite{YI2016105, wang2023dynamic}, Transient Grating
Spectroscopy (TGS) \cite{TGS} etc.  have recorded important microstructural
properties on irradiation such as defect concentration, dislocation densities,
swelling etc. at various dpa including lower dpa that can be used to match
against the simulation studies. \\

In this paper we carry out MD simulations of successive collision cascades (SCC)
using two different interatomic potentials and at two different PKA energies. We
investigate different defect properties as a function of DPA reaching upto
0.1 dpa for 20 keV and 0.2 dpa for 50 keV collision cascades. In the next
section we describe the simulations and methods. This is followed by the
sections describing our results and discussing them further. Finally the
conclusions are presented.

\section{Methods}
\subsection{MD Simulations of Successive Collision Cascades}
\label{DescMD}
MD simulations of successive collision cascades were carried out at PKA energies 
of 20 keV and 50 keV in a simulation domain consisting of $100 \times{} 100 
\times 100$ unit cells of tungsten using two different IAP. One of the 
potentials is an embedded atom method (EAM) based potential called the DND-BN 
potential \cite{DND-BN}, and the other is a machine learning (ML) based quantum 
accurate potential called the SNAP potential \cite{wood2017SNAPPot}. Both the 
potentials take care that the pair interactions when the atoms in the solid 
come close to each other are treated by the universal ZBL potential 
\cite{ZBLPot}. The simulation region is first equilibrated at zero pressure and 
300 K temperature using a NPT ensemble, with periodic boundary conditions (PBC) 
along all the three directions. After equilibration, a randomly selected PKA is 
launched in a random direction. The random directions are chosen by identifying 
uniformly distributed points on a random sphere centered at the position of the 
PKA. PBC are used for the collision cascade simulations too with a varying 
time-step such that the fastest atom in the system moves less than 0.1 \AA. 
Electronic stopping is included as a frictional term calculated by the 
Lindhard-Scharff model \cite{HarshEStopLAMMPS}. The collision cascade 
simulations are carried out for 20 ps using an NVE ensemble. Each cascade 
simulation is followed by an NPT simulation at zero pressure and 300 K 
temperature for 10 ps to relax the damage. Five trials of successive collision 
cascade simulations were carried out at each energy and for each potential for
statistics. The dpa is calculated using the standard NRT equation \cite{WasRadDamIonBeams}
with threshold displacement energy of 70 eV \cite{TGS}.

\subsection{Analysis of the micro-structure damage}

The parallelized version of CSaransh \cite{BhardwajJOSS}, a software suite to
analyze large database of MD simulations of collision cascades, was used to
analyze the simulation results. It outputs various defect properties such as
the number of defects, defects in clusters, the defect morphology and size distributions.
We analyze the simulation box after every collision cascade for all the five trials which
makes a total of 45000 simulation frames to analyze.

A number of factors make analysing high dose SCC simulations particularly more
challenging compared to the single PKA simulations. The boundaries in current
SCC simulations are not fixed which may result in the movement of whole lattice
structure. Due to this global shift in positions the methods that use initial
lattice positions as template to find defects such as Wigner-Sietz become
infeasible. The presence of large
defects sometimes wrapping across the boundary makes it even more difficult to
estimate the absolute perfect lattice positions to compare with. The defect
analysis algorithms of Csaransh named as AnuVikar use specific minimization
methods to make this estimation efficient as well as accurate at such a large
scale of analysis. The calculation of different defect morphologies and 
properties associated with them such as the total defects and surface defect
atoms in a dislocation loop are not directly available from the existing
algorithms.

The Csaransh software suite used in our study uses, Anuvikar algorithm, a modification
of W-S and ES methods \cite{BhardwajClassify} to find defects. The algorithm
does not take an initial set of positions as input template in case of known
crystal structures such as bcc, fcc, hcp etc. The initial position of the
beginning of lattice is estimated by optimizing on all the atomic positions.
This makes it work well for successive collision cascades where the lattice can
shift as a whole and there is a possibility of volumetric swelling. Each atomic
coordinate is then associated with its nearest lattice site using modular
arithmetic which is more efficient than making geometric structures or neighbour
lists as done in W-S method \cite{BhardwajClassify}. The lattice sites that are
not marked as the nearest lattice site by any atom are vacancies and the atoms
that are associated with over-occupied lattice sites are SIAs (self-interstitial
atoms) or displaced atoms. We use SaVi, a computational graph based algorithm
\cite{BhardwajClassify, BhardwajSavi} to classify defect morphologies. It
characterizes the orientation and internal morphology of the defects into
various classes using computational graphs and computational geometry. The
detailed analysis allows calculating the surface defects in a dislocation loop
which we use for experimental comparison. It has recently been parallelized for
quick analysis of a large database of MD simulations using C++ and python and
can typically carry out the analysis of thousands of MD collision cascade atomic
positions output in less than an hour and outputs JSON (JavaScript Object
Notation) files for further specific analysis. 

\section{Results}
We perform subsequent collision cascade simulations at two energies viz. 20 keV
and 50 keV, reaching at the approximate dpa levels of 0.1 and 0.2, respectively.
Table.\ref{DPATable} shows the dpa obtained by the MD simulations for each of
the PKA energies and IAP explored. For each case we have performed five sample
runs for statistics. Most of the plots in the results section show mean of these
five samples along with error bars.
\begin{table}[ht]
\centering
\caption{The dpa obtained from the successive MD simulations of collision
cascades}
\label{DPATable}
\begin{tabular}{|c|c|c|c|} \hline
PKA Energy & IAP used & Number of    & dpa \\
 (keV)     &          & PKA Launched &     \\ \hline
20         & DND-BN   & 2500         & 0.11 \\ \hline
20         & SNAP     & 2500         & 0.11 \\ \hline
50         & DND-BN   & 2000         & 0.22 \\ \hline
50         & SNAP     & 2000         & 0.22 \\ \hline
\end{tabular}  
\end{table}

\subsection{Defect Count}

\begin{figure}[H]
\centering{\includegraphics[width=\textwidth]{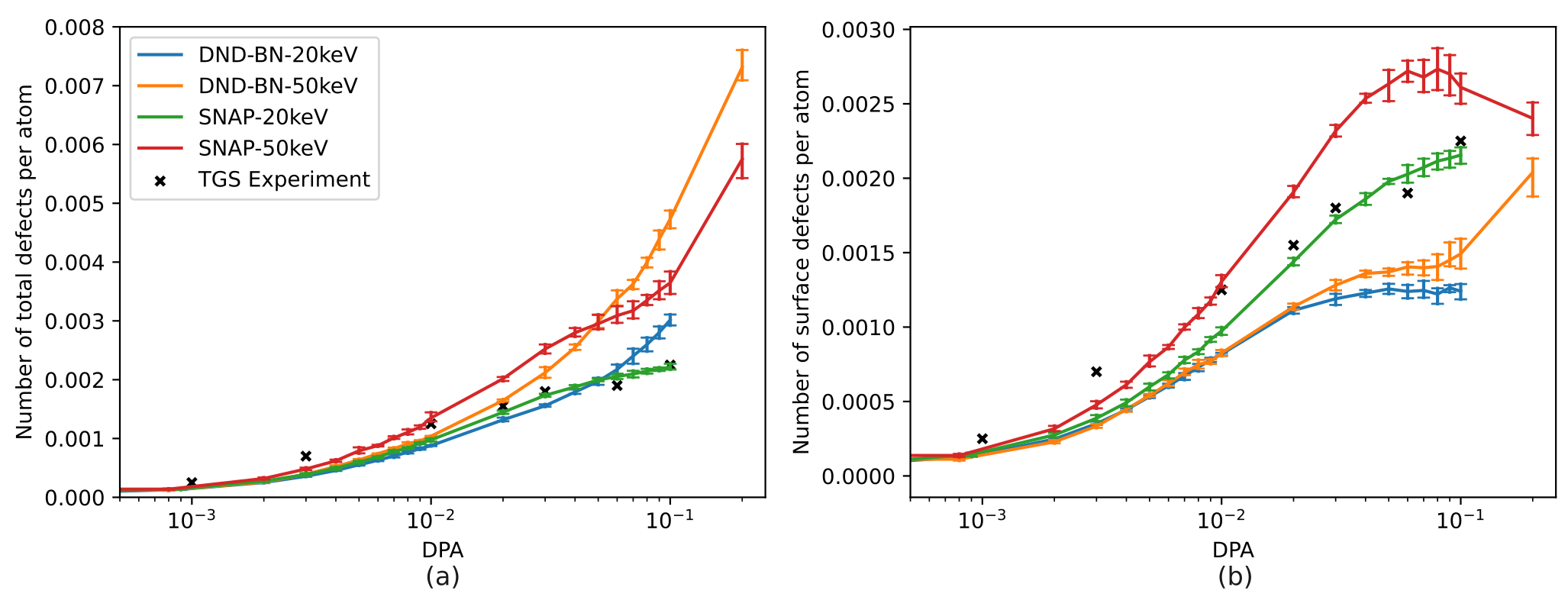}}
\caption{\label{Fig1} At various dpa, the (a) number of total point defects per atom
  (b) number of single point defects and defects on the surface of a cluster 
  for the four cases specified in Table.\ref{DPATable}.  
}
\end{figure}

Fig.\ref{Fig1} (a) shows the variation of the number of defects 
per atom as a function of dpa. Fig. \ref{Fig1} (b) includes the single point defects and
only the surface atoms of an interstitial defect cluster. It is
same as counting the point defects at the boundary or around the dislocation loop of
an edge dislocation and ignoring the point defects that are inside. The atoms on the
surface are known to have higher-strain than the atoms inside the loop
\cite{BhardwajSavi}. Many of the experiments, and material properties are known to
correlate well with the surface atoms of the defect
rather than the total number of point defects. We can see that when only the
surface atoms are counted different potentials and energies show similar values
and trends. The simulations that produce bigger defect clusters observe a
decrease in the defects when only the surface defects are counted.  For
this reason, the values for DND-BN potential see more reduction than the SNAP
potential. The SNAP potential at 20 keV gets least affected as it has small
defects and very few edge dislocation loops. We will see this in detail in
later sections on SIA defect size distribution and
morphology details.

Fig.\ref{Fig1} also shows the number of defects estimated by the transient grating spectroscopy \cite{TGS}
experiment for the 20 MeV self-ion irradiation of tungsten. The Transient Grating
Spectroscopy experiment is indicative of the number of surface defects rather than the
total defects \cite{TGS}. All the simulations show a decent match with the experimental
results. The SNAP potential at 20 keV shows the best match with the experiment.
The DND-BN potential shows slightly lower number of surface defects. However, both
the potentials seems to be approaching the experimental values at 0.2 dpa.

We find that the number of total defects at different dpa differ
based on energy of the PKA but the total number of surface defects show similar
trends \ref{Fig1}. Even if the surface defects rise at certain point they start to decrease
as the defect clusters grow in size and start to merge together. The saturation
point of 0.002 defects per atom matches that of the experiments \cite{TGS}. The
total defects deviate largely based on the IAP used, however the surface defects
again seems to be converging as the dpa increases. The experiment is conducted
at 20 MeV. A very high energy PKA results in a combination of subcascades of
lower energy.

\begin{figure}
\centering{\includegraphics[width=\textwidth]{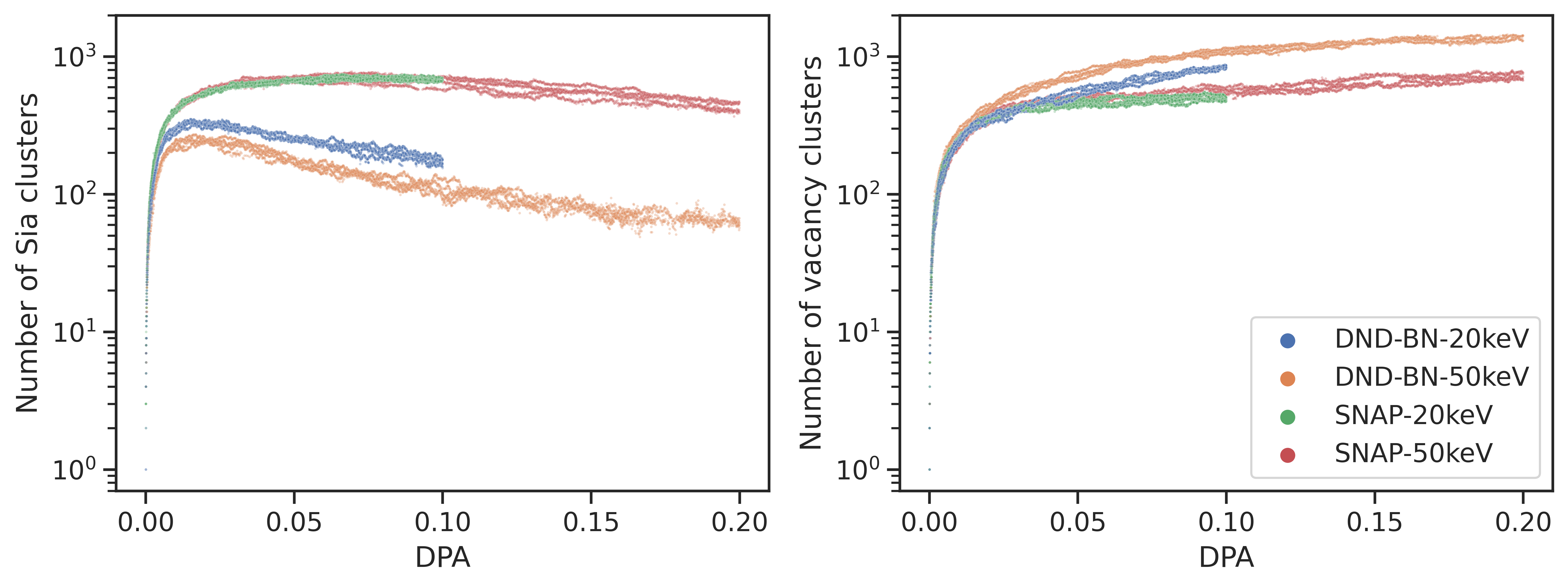}}
\caption{\label{Fig2} Variation of the number of defect clusters of (a) SIA and (b) Vacancy, with dpa for 
the four cases specified in Table.\ref{DPATable}}
\end{figure}

Fig.\ref{Fig2} shows the variation in the number of SIA  and vacancy 
clusters as a function of dpa. The number of SIA clusters for DND-BN start
to decrease while for SNAP the values only saturate. The decrease in clusters in
DND-BN is due to the big SIA clusters in DND-BN which start to merge together with
the increase in defect density. There is a quantitative separation based on the
IAP rather than the PKA energy as was seen in the case of number of defects. 

\begin{figure}[H]
\centering{\includegraphics[width=.6\textwidth]{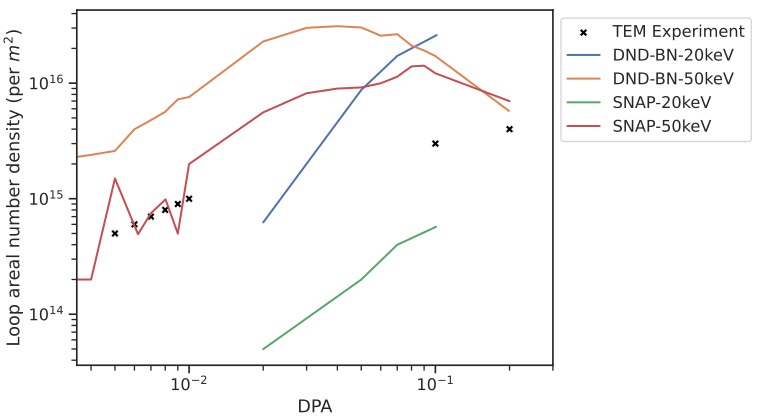}}
\caption{\label{Fig3} Loop density of TEM-visible defects as a function of
dpa for the four cases specified in Table.\ref{DPATable}} 
\end{figure}

Fig.\ref{Fig3} shows the number density of dislocations greater than 1.5 nm
(or having defect count greater than 30) which are big enough to be observed in
a TEM \cite{YI2016105}. The defect density of such defects in SNAP at 20 keV
remains lower compared to others. In other cases the defects increase with dose
and then start to decrease as the defect clusters grow in size and start to
merge together.  Since, biggest defects are formed in DND-BN simulation at 50
keV, the saturation appears earliest (at around 0.4 dpa) in it. The
experimental values from a TEM experiment performed by 150keV self ion
irradiation \cite{YI2016105} are also shown. The values for SNAP potential at 50
keV perfectly matches initially and again appear to be approaching the
experimental value at 0.2 dpa with an over-estimation in the mid dpa range. The
values rise till around 0.1 dpa and then start to decrease. The decrease in 
loop density is due to merging of defects to form dislocation strings and
networks. The decrease of loop density starting at around 0.1 dpa and then saturating
by 0.2 dpa at a much lower value agrees well with the recent TEM experiment \cite{wang2023dynamic}. 
The value for loop density significantly differs for 20
keV and 50 keV. The experiment values in the plot are for 150keV self ion-irradiation which is close to the subcascade
threshold of W \cite{BhardwajSubcascades}. A 150 keV cascade
would break into combination of lower energy sub-cascades in different
proportions \cite{BhardwajSubcascades}.  In such a case, the SNAP potential's
slightly lesser values of loop density at 20 keV and slightly higher values at
50 keV compared to the experiment seems to be balancing out for a better
agreement than the consistently higher estimations of DND-BN at both the
energies.

\subsection{In-cluster defects}

\begin{figure}[H]
\centering{\includegraphics[width=\textwidth]{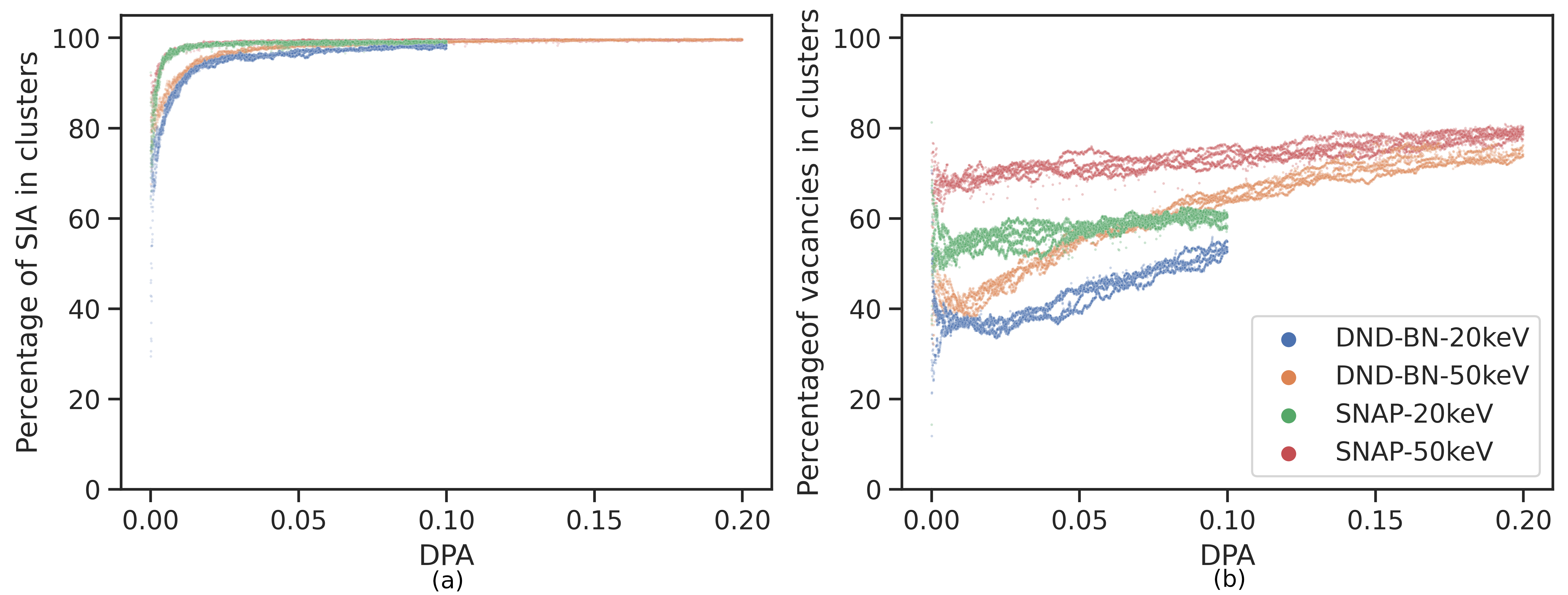}}
\caption{\label{Fig4} Variation of the number of in-cluster (a) SIAs and (b) Vacancies with
dpa for the four cases specified in Table.\ref{DPATable}}
\end{figure}

Fig.\ref{Fig4} shows in-cluster SIAs and vacancies as a function of dpa. From
Fig.\ref{Fig4} (a) it can be seen that the percentage of interstitials in
clusters quickly rises and reaches a value greater than 95 \% before 0.01 dpa for SNAP and
by 0.04 dpa for DND-BN. The rising curves are separated by IAP type rather
than the PKA energy. For the in-cluster vacancies case shown in Fig.\ref{Fig4}
(b), there is a gradual rise with dpa. Unlike the in-cluster interstitials case,
there is a well defined separation in the variation of the percentage of
vacancies in clusters for both the PKA energies and for the IAP.  The vacancies
also show greater variability for different simulation trials.  In general the
SNAP potential shows a larger number of in-cluster vacancies. \\

\subsection{Vacancy Size Distribution}

\begin{figure}[ht]
\centering{\includegraphics[width=\textwidth]{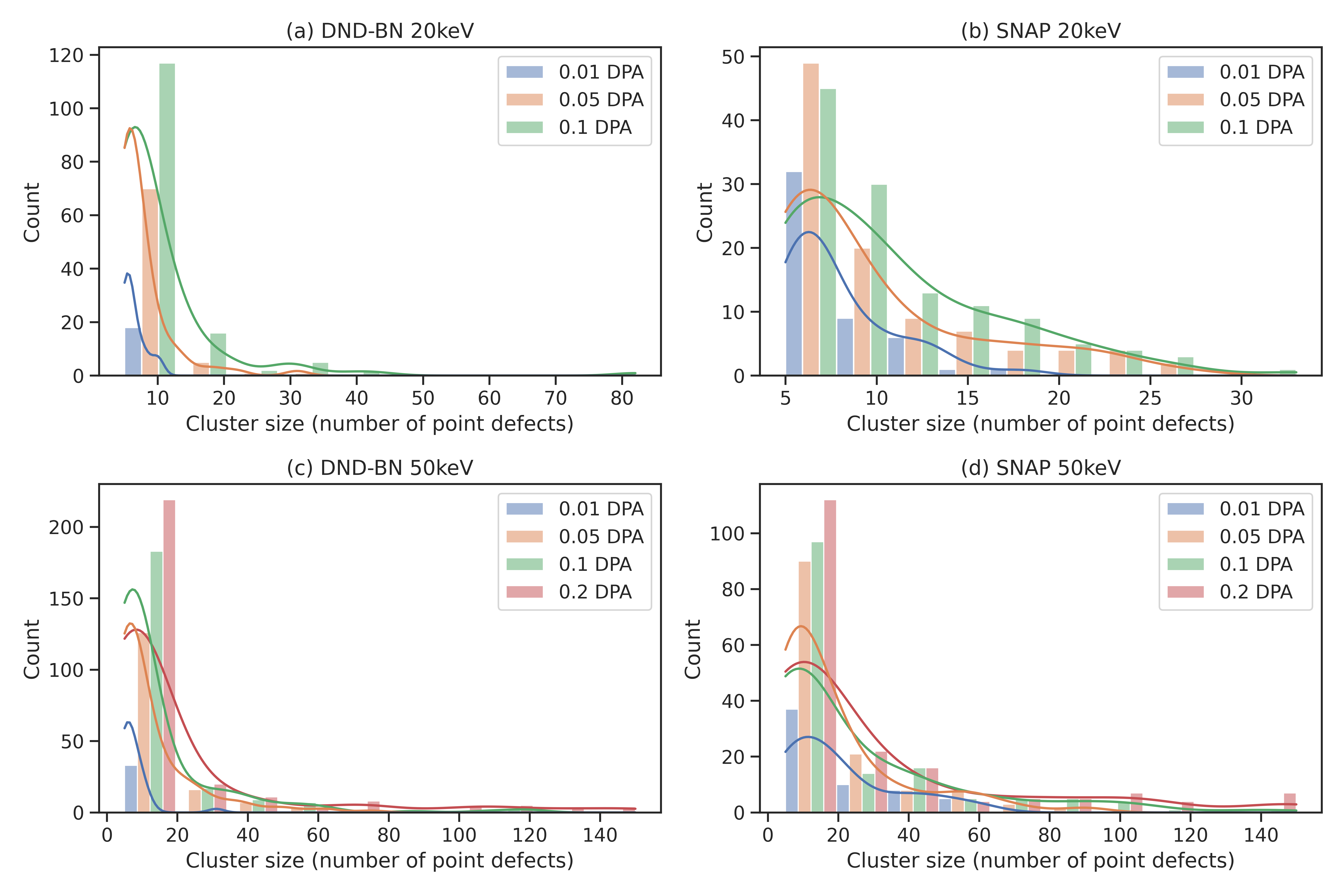}}
\caption{\label{Fig5} Vacancy cluster size distribution at 0.01, 0.05 and 0.1 dpa
for the four cases specified in Table.\ref{DPATable}}
\end{figure}

Fig.\ref{Fig5} shows the vacancy cluster size distribution for all the vacancy clusters
with size more than 5 number of vacancies. As the dpa increases
the maximum vacancy size and the counts for all the sizes both grow
in all the cases. The vacancy sizes for the same dpa at different energies are vastly different.
The 50keV energy cases for both the IAPs have bigger vacancy clusters compared to 20keV.
This is expected because at both these energies, W still prefers single sub-cascade
\cite{BhardwajSubcascades} and the 50keV single sub-cascade will have a bigger central core
of vacant sites. It can be seen that the distributions of SNAP IAP have slightly
fatter tails signifying tendency to have more of bigger vacancy clusters. This
is contrary to the SIA cluster size distribution as we will show in the following
results.

\subsection{Defect Morphologies}

\begin{figure}[H]
\centering{\includegraphics[width=\textwidth]{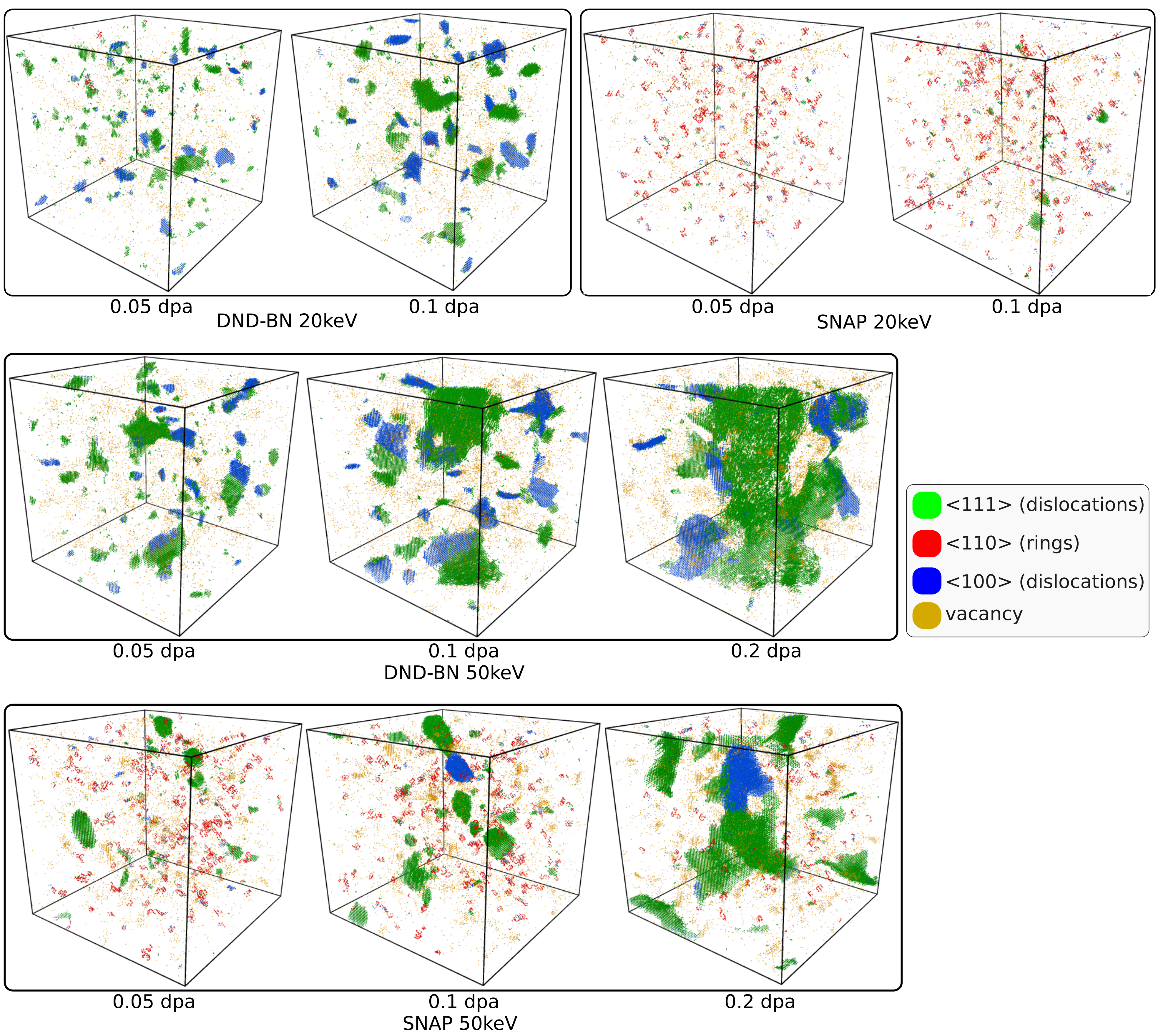}}
\caption{\label{Fig8} SIA defects at different dpa in the simulation box. The colours
  show the SIA dumbbell/crowdion orientation. The bigger clusters of \hkl<1 1 1> and \hkl<1 0 0> 
  dumbbells/crowdions form the edge dislocations with the respective burgers vector. 
  Almost all the \hkl<1 1 0>
  dumbbells/crowdions form C15 like rings. For the same dpa, the density and sizes of 
  defects at higher energy are more. SNAP shows high number of rings while bigger dislocations
dominate in DND-BN.}
\end{figure}

\begin{figure}[ht]
\centering{\includegraphics[width=.3\linewidth]{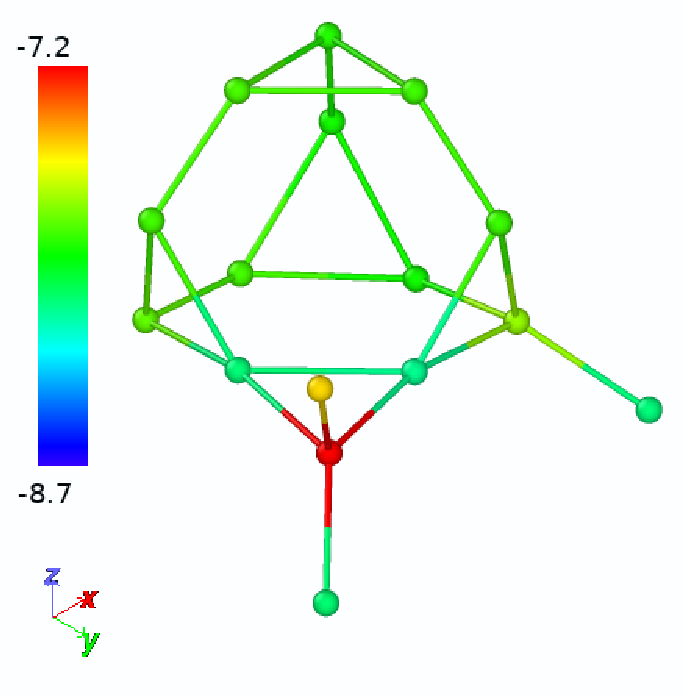}}
\caption{\label{Fig9} A typical C15 ring (denoted by $@$) like structure observed in
  tungsten. The color-map represents the potential energy of the atoms (in eV) - visible
in color in the digital version of this article.}
\end{figure}

Fig.\ref{Fig8} shows the SIA defects in the simulation box at different dpa.
The SIA are coloured based on the orientation of dumbbells/crowdions as found
using SaVi algorithm \cite{BhardwajSavi}. The bigger clusters of parallel
\hkl<1 1 1> and \hkl<1 0 0> dumbbells/crowdions form the edge dislocations with
burgers vector of 1/2\hkl<1 1 1> and \hkl<1 0 0>, respectively. The \hkl<1 1 0>
dumbbells sometimes appear at the interface between a multi-component
dislocation (a cluster of two of more dislocations) but mostly they form the C15
like rings or their basis \cite{BhardwajSavi}. Fig.\ref{Fig9} shows a typical
C15 3d ring defect. A defect can be composed of multiple such ring units. The
frames in the Fig.\ref{Fig8}, qualitatively show the differences in morphology
of SIAs in the four cases. The SNAP IAP forms lots of small C15 like rings while
in case of DND-BN we observe bigger \hkl<1 1 1> and \hkl<1 0 0> defects being
formed as the dpa increases. The presence of rings in SNAP can be noted at high
dpa as well.

The number of \hkl<1 0 0> dislocations increase in DND-BN. The \hkl<1 0 0 >
dislocations can get formed by the interaction of two non-collinear 
dislocations \cite{YI2016105}. The
diffusion and creation of more dislocations results in dislocations being
combined to form bigger multi-dislocations and dislocation network
\cite{wang2023dynamic}. This results in reduction of the number of surface defects
and the number of clusters as shown earlier in Fig.\ref{Fig1} and
Fig.\ref{Fig3}. The sizes of the defects for 20 keV at 0.1 dpa match well with that of 50
keV at 0.05 dpa for the same potential. This is due to the
bigger clusters being formed right after a 50 keV collision cascade as compared
to a 20 keV collision cascade. This supports the variation of loop density of TEM-visible
loops for the two energies in Fig.\ref{Fig3}.

\begin{figure}[H]
\centering{\includegraphics[width=\textwidth]{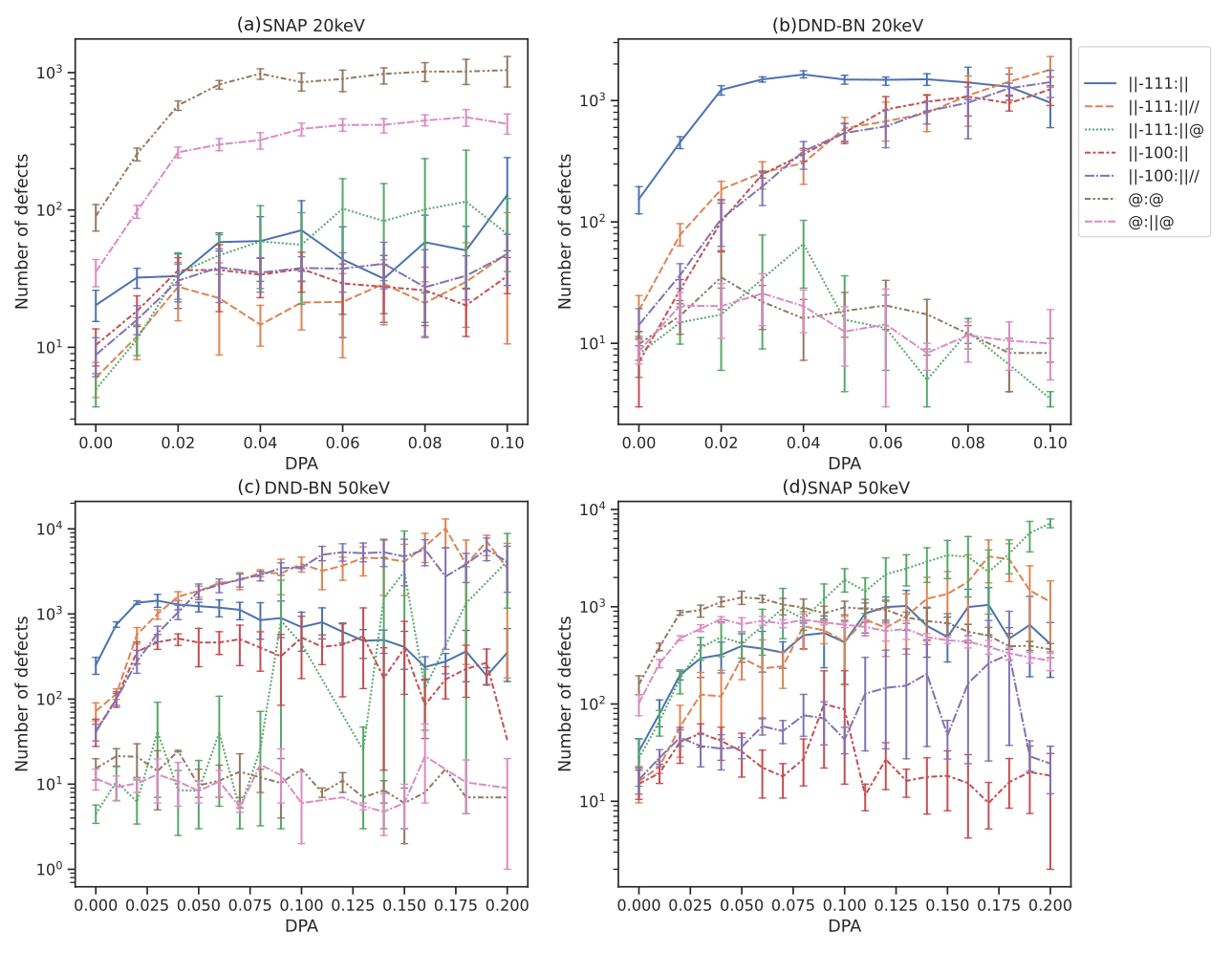}}
\caption{\label{Fig6} Variation in the number of defect clusters of various
types with dpa for the four cases specified in Table.\ref{DPATable}. 
The symbols in the legend represent morphology of single dislocation $||$ with \hkl<1 0 0> or \hkl<1 1 1> Burgers vector or a ring $@$. These may also appear in mixed defect configurations such as dislocation-ring ($||@$) or multi-dislocation ($||//$). }
\end{figure}

\begin{figure}[H]
\centering{\includegraphics[width=\textwidth]{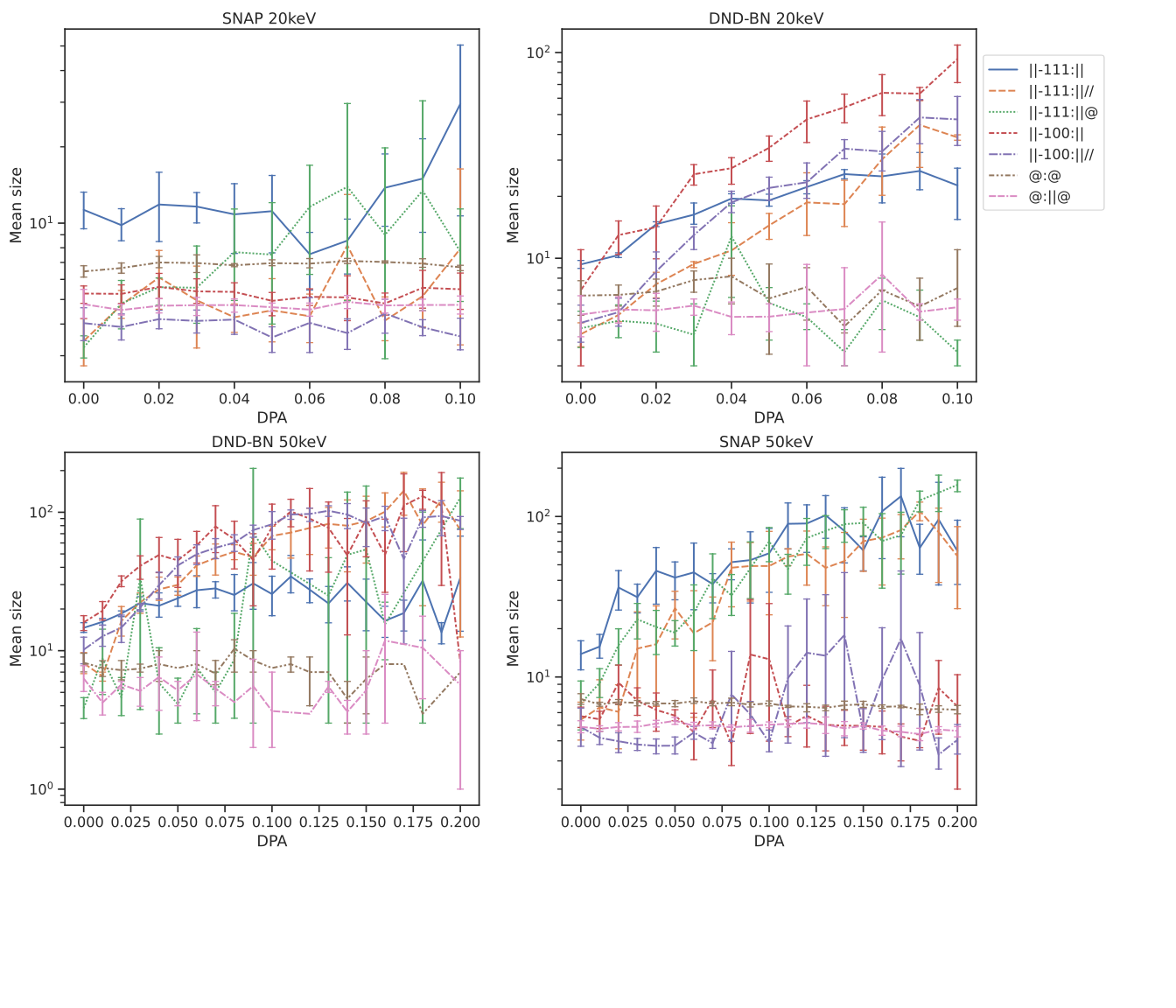}}
\caption{\label{Fig7} Variation in the mean size of SIA defect clusters of various
types with dpa for the four cases specified in Table.\ref{DPATable}. Here the
size refers to the number of SIA constituting a defect cluster.
The symbols in the legend represent morphology of single dislocation $||$ with \hkl<1 0 0> or \hkl<1 1 1> Burgers vector or a ring $@$. These may also appear in mixed defect configurations such as dislocation-ring ($||@$) or 
multi-dislocation ($||//$). }
\end{figure}

Fig.\ref{Fig6} and Fig.\ref{Fig7} quantify the above observations, showing
the distribution of different morphologies and their mean sizes, respectively.
We have resolved the different morphological components in a defect composed
of multiple dislocations or dislocation and ring component. Different
morphologies are represented by different ASCII letters for the figure labels.
$||$ signifies clusters with dumbbells/crowdions arranged in parallel
configuration which form 1/2 \hkl<1 1 1> and \hkl<1 0 0> edge dislocations in W.
The symbol $@$ signifies C-15 like rings and its basis configurations. There are
also mixed morphologies defects with multiple components. We distinguish between
dislocations that are part of a mixed dislocation defect (cluster having
multiple loops) and defects that are composed of homogeneously one dislocation
loop. The symbol $||//$ is for defects composed of two or more dislocations in
different orientations. In the Figure legend, the symbol $||-111:||//$
represents the number of defects that form \hkl<1 1 1> edge dislocation and are
part of a multi-component dislocation while $||-111:||$ is the number of defects
that form pure \hkl<1 1 1> dislocation. The symbol $||-111:||@$ is for the
configuration that form \hkl<1 1 1> dislocation while also being attached to a
ring $@$. The properties such as diffusion, stability and interaction of the
same component $||-111$ can vary in the above three cases.\\

The number of defects in different morphologies remain similar for the two
energies of same IAP, whereas notable difference exists across the two IAPs.
Fig.\ref{Fig6} shows that the SNAP IAP results in significantly higher number of
ring like structures. The rings remain high through out the DPA range.  The
DND-BN potential in contrast shows very few ring like structures and the
micro-structure is dominated by the edge dislocations $||$. Initially, the 
1/2\hkl<111> edge-dislocations in pure ($||$) as well as mixed $||//$ defects
are highest however at higher dpa the \hkl<1 0 0> dislocations become dominant
and pure \hkl<1 1 1> dislocations start to drop. In contrast, SNAP IAP shows
very less \hkl<1 0 0> edge-dislocations.\\

The mean sizes of the SIA defect clusters in various morphologies as a function
of dpa in fig.\ref{Fig7} shows that the pure \hkl<1 1 1> parallel defects ($||-111:||$)
that form glissile 1/2 \hkl<1 1 1> edge dislocations have a mean size of few
tens of defects in all the four cases. The mean size of pure as well as mixed
\hkl<1 0 0> dislocation component is slightly higher than \hkl<1 1 1> in DND-BN.
In SNAP $||-100:||$ rarely gets formed. The small \hkl<1 0 0> edge dislocations
defects are not stable in W \cite{Bhardwaj100Stability} as a result the mean
size of \hkl<1 0 0> dislocations ($||-100$) is higher in DND-BN than the \hkl<1
1 1> parallel defects which includes bigger \hkl<1 1 1> edge dislocations as
well as many small parallel pairs, triplets etc. of dumbbells which form very
often with every collision cascade. \\

For SNAP, the mean size of rings remain less that 10 although the maximum is a
few tens of point defects. A good number of bigger \hkl<1 1 1> edge
dislocations appear in conjunction with a ring component ($||-111:||@$). It is
seen that for both the IAPs the C15 ring like structures ($@$ and $@||$) are
small while the \hkl<1 1 1> and \hkl<1 0 0> defects are relatively much bigger,
especially for the DND-BN potential. This matches the trend of larger sized
loop formation at higher dpa as seen in experiments \cite{RayaproluProtonW, wang2023dynamic}. \\

The trend of mean sizes for dislocations ($||$) are slightly different for the two
energies. For DND-BN at 20keV, the sizes of \hkl<1 0 0> and \hkl<1 1 1> in multi-component dislocation defect ($||//$)
start at lower value and increase gradually in size with DPA while
at 50 keV the sizes of these morphologies get comparatively higher from the
beginning. For SNAP, at 20keV the mean sizes remain slightly lower 
when compared with 50keV. 

\subsection{Swelling}

\begin{figure}[H]
\centering{\includegraphics[width=.7\textwidth]{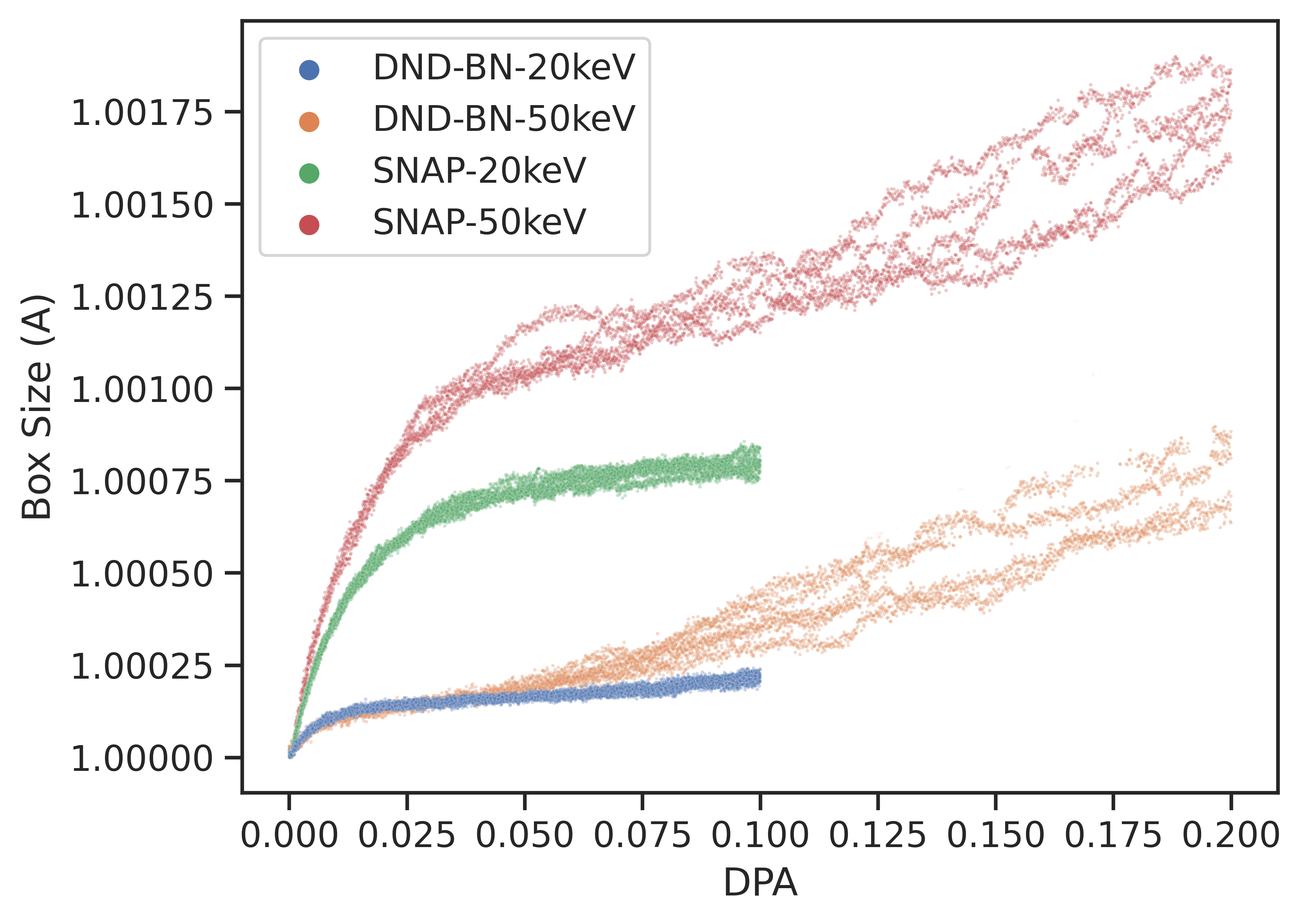}}
\caption{\label{Fig10} Fractional increase in the simulation box size as a
function of dpa for both the potentials and both the energies.}
\end{figure}

Fig.\ref{Fig10} shows the expansion or swelling of the simulation box as a
function of dpa. The increase in simulation box size is more significant in
SNAP IAP than DND-BN. The swelling in 50 keV is more for both the
potentials. The experimental values for swelling at different doses and temperatures is 
reported to be in the range of 0.2 to 0.4\% \cite{el2018loop,nguyen2021first}.
The swelling suggested by DND-BN is quite low while SNAP potential especially at 50keV
is closer to the experimental results.

\section{Discussion \label{Discuss}}

The SNAP and DND-BN potentials that we have used in our study show significant
differences in defect morphologies and sizes. The SNAP machine learning potential 
which can be considered quantum accurate, produces high number of smaller, stable and sessile 
ring defect clusters in addition to \hkl<1 1 1> dislocation loops while DND-BN potential
mainly produces big \hkl<1 1 1> as well as \hkl<1 0 0> dislocations. Despite these differences
when we look at the damage properties that can be measured
and are known to correlate well with the material properties, such as the number
density of surface point defects and the number of TEM-visible defect loops, the
two potentials show similar trends (as shown in Fig.\ref{Fig1} and Fig.\ref{Fig3}). 
The SNAP potential does show a slightly better match with the experimental
values of surface defect density and swelling which suggests that its prediction
of high number of stable rings is likely to be accurate. Since, these rings are
relatively small, their signatures need to be exclusively looked for in
experiments. A high number of sessile rings can restrict the diffusion of mobile
defects. An interesting investigation would be to observe how the defects with
different mobility will interact with each other and affect the microstructural
properties with time evolution. 

The irradiation induced damage in materials is quantified by the number of
displacements per atom (dpa) since many of the material properties are known to
correlate well with the dpa irrespective of the energy of the incident
particle \cite{WasRadDamIonBeams}. The total number of surface defects show
similar trends with dpa for both the PKA energies and does not show significant
deviation based on the IAP. The number of surface defects are rarely noted in
the literature when comparing collision cascades carried out with different
potentials or materials. The current study suggests that the number of surface
defects is an important parameter for such studies. 

% The number density of defect clusters and in-cluster SIAs is same for both the
% PKA energies while the SIA and vacancy cluster sizes are notably high at
% 50 keV PKA (Fig.\ref{Fig5} and Fig.\ref{Fig7}). This energy dependence of
% cluster sizes also makes the loop number density of TEM-visible defects 
% differ for the two energies Fig.\ref{Fig3}. 

\section{Conclusions}

MD simulations of successive collision cascades have been carried out in 
tungsten at PKA energies of 20 keV and 50 keV using an EAM and a ML based 
inter-atomic potential. For both the potentials, five sample simulations were
carried out with successive PKAs reaching a dose of 0.1 and 0.2 dpa for 20keV
and 50keV, respectively. Various properties such as  the
the number density of defect clusters and in-cluster SIAs is same for both the
PKA energies while the SIA and vacancy cluster sizes are notably high at
higer energy (Fig.\ref{Fig5} and Fig.\ref{Fig7}). The number of surface defects
are observed to saturate at the same level as experiments irrespective of energy
of the PKA or the potential used.

The defect morphology of the two potentials are significantly different. The SNAP
potential produces smaller, stable and sessile C15-like ring defect clusters in addition
to \hkl<1 1 1> dislocation loops while the EAM potential produces bigger \hkl<1 0 0>
and \hkl<1 1 1> dislocations. The SNAP potential which is considered more closer
to ab-initio calculations also shows a better overall match with the experimental
values of the swelling, number of surface defects and
number density of TEM visible defects. Very high number of \hkl<1 0 0>
dislocation loops suggested in DND-BN potential at high dpa also appears to be
an over-estimation compared to the current experiments \cite{YI2016105}. The
predictions of the SNAP potential for a high number of stable rings is very
likely to have practical considerations and the existence of such rings must be
investigated further. The presence of small sessile defects can have significant
impact on defect evolution and material properties. 

\section*{Acknowledgements}
We acknowledge the HPC team at CAD for the HPC maintenance and for
support in installing LAMMPS. 

\section*{References}
\bibliographystyle{elsarticle-num}
\bibliography{WMDDPAStudies.bib}

\end{document}